\documentclass{article}

\PassOptionsToPackage{numbers,compress}{natbib}
\usepackage{graphicx}
\usepackage{booktabs,tabularx}
\usepackage{wrapfig}
\usepackage{fix-cm}

 \usepackage[preprint, creativeai]{neurips_2025}
\def\eg{\emph{e.g.}}

\newcommand\myfootnotestyle[1]{\ifcase#1 \or \ding{182}\or \ding{183}\or
\ding{184}\or \ding{185}\or \ding{186}\or \ding{187}%
\or \ding{188}\or \ding{189}\or \ding{190}\or \ding{191}\else *\fi\relax}

\usepackage[utf8]{inputenc} 
\usepackage[T1]{fontenc}    
\usepackage{hyperref}       
\usepackage{url}            
\usepackage{booktabs}       
\usepackage{amsfonts}       
\usepackage{nicefrac}       
\usepackage{microtype}      
\usepackage{xcolor}         
\usepackage{etoolbox} 
\newcommand\chapquoteauthor{} 
{%
  \renewcommand\chapquoteauthor{#1}%
  \begin{list}{}{\leftmargin=2em \rightmargin=2em}
  \item\relax\itshape
}
{%
  \par\smallskip
  \ifstrempty{\chapquoteauthor}{}{\hfill\upshape--- \chapquoteauthor}%
  \end{list}
}

\title{Uncovering Strategic Egoism Behaviors in Large \\Language Models}

\newcommand*{\affaddr}[1]{#1} 
\newcommand*{\affmark}[1][*]{\textsuperscript{#1}}

\author{
    Yaoyuan Zhang\affmark[1],
    \textbf{Aishan Liu\affmark[1]}, 
    \textbf{Zonghao Ying\affmark[1]\thanks{Corresponding Author},}  
    \textbf{Xianglong Liu\affmark[1,2,3]}
    \textbf{Jiangfan Liu\affmark[1]}, \\
    \textbf{Yisong Xiao\affmark[1]}, 
    \textbf{Qihang Zhang\affmark[4]}, 
    \\
    \affaddr{\affmark[1]SKLCCSE, Beihang University}
    \affaddr{\affmark[2]{Zhongguancun Laboratory}}\\
    \affaddr{\affmark[3]{Institute of Dataspace}}
    \affaddr{\affmark[4]{Beijing Jiaotong University}}
}

\begin{document}

\maketitle

\begin{abstract}

Large language models (LLMs) face growing trustworthiness concerns (\eg, deception), which hinder their safe deployment in high-stakes decision-making scenarios. In this paper, we present the first systematic investigation of strategic egoism (SE), a form of rule-bounded self-interest in which models pursue short-term or self-serving gains while disregarding collective welfare and ethical considerations. To quantitatively assess this phenomenon, we introduce SEBench, a benchmark comprising 160 scenarios across five domains. Each scenario features a single-role decision-making context, with psychologically grounded choice sets designed to elicit self-serving behaviors. These behavior-driven tasks assess egoistic tendencies along six dimensions, such as manipulation, rule circumvention, and self-interest prioritization. Building on this, we conduct extensive experiments across 5 open-sourced and 2 commercial LLMs, where we observe that strategic egoism emerges universally across models. Surprisingly, we found a positive correlation between egoistic tendencies and toxic language behaviors, suggesting that strategic egoism may underlie broader misalignment risks.

\end{abstract}

\section{Introduction}

LLMs are increasingly deployed in high-stakes decision-making domains such as healthcare, finance, and public administration. Existing safety strategies primarily focus on surface-level harms such as toxicity~\citep{liang2023badclip,liang2025revisiting,liu2024compromising,xiao2024bdefects4nn,liang2024red,liu2025elba,ying2024safebench}, social bias~\citep{xiao2024genderbias,xiao2025fairness}, jailbreak attacks~\citep{li2024semantic,ying2024jailbreak,jing2025cogmorph,ying2025reasoning,wang2025manipulating,ying2025pushing,liu2025agentsafe,ying2024unveiling,ying2025towards}, and lack of truthfulness~\citep{chen2024less,chen2024interpreting,chen2025less,ho2024novo}, but often overlook covert self-interested behaviors that may emerge when models face reward pressure in decision-making contexts \citep{rtp2020,Nangia2020CrowSPairs,Lin2022TruthfulQA,xiao2025fairness,ying2025reasoning,ying2025securewebarena}. For example, models may engage in unfair resource allocation to maximize personal gain, or selectively withhold information to preserve advantages. Emerging evidence shows that LLMs’ deceptive and manipulative behaviors reflect dark-personality tendencies, posing significant safety risks \citep{williams2024targeted,Scheurer2024Deceive}.
Moreover, recent Cambridge-affiliated research shows that most alignment and prompt-engineering approaches remain concentrated at the linguistic level rather than the behavioral level, lacking analyses that characterize model ``personality'' from observed behavior \citep{Han2025PersonalityIllusion}. These results underscore the need for a behaviorally grounded alignment perspective that analyzes LLM behavior to address latent self-interested tendencies currently missed by safety benchmarks.

In this paper, we present the first systematic investigation of Strategic Egoism (SE). We define SE as a decision-making tendency in which agents prioritize personal or short-term rewards under explicit procedural constraints while disregarding social responsibility and the welfare of others. To make SE measurable, we introduce SEBench, a benchmark comprising 160 single-role decision scenarios across five domains (education, markets, government, enterprise, and healthcare). Each scenario specifies explicit procedural constraints and role-specific incentives (\eg, promotion, performance review) and leverages personality psychology (\eg, the Dark Triad) to construct choice sets that quantify six self-interest tendencies (\eg, manipulation, unfair allocation). Regarding the design of options, they are contextualized to the scenario and expressed as actionable behaviors, such as spreading misinformation to steer others' decisions or inflating contributions while concealing defects to secure greater personal gain. The implementation of these two behaviors relies on deception and aligns with Machiavellian tendencies in personality psychology. Combining psychological theories yields clear behavioral measures that help us capture incentive-driven harmfulness often missed by common metrics. Ultimately, we found that across seven mainstream LLMs, egoistic behaviors account for an average of 69.11\%, highlighting the urgent need to improve decision-making safety mechanisms under incentive temptations. Our main contributions are summarized as follows:
\begin{itemize}
\item We formalize \textit{Strategic Egoism} and release \textit{SEBench}: 160 single-role scenarios across five domains with explicit rules, role-specific incentives, and six egoistic dimensions plus a compliant option.
\item We introduce the \textit{SE Rate (SER)} and per-type distributions for interpretable, fine-grained measurement of egoism.
\item Using \textit{SEBench}, we evaluated 7 LLMs and found pervasive SE behaviors, with manipulation, coercion, and oversight evasion as the most common strategies; moreover, egoistic tendencies were positively correlated with toxicity.
\end{itemize}
\begin{figure}[t]
  \centering
  \includegraphics[width=1.0\linewidth]{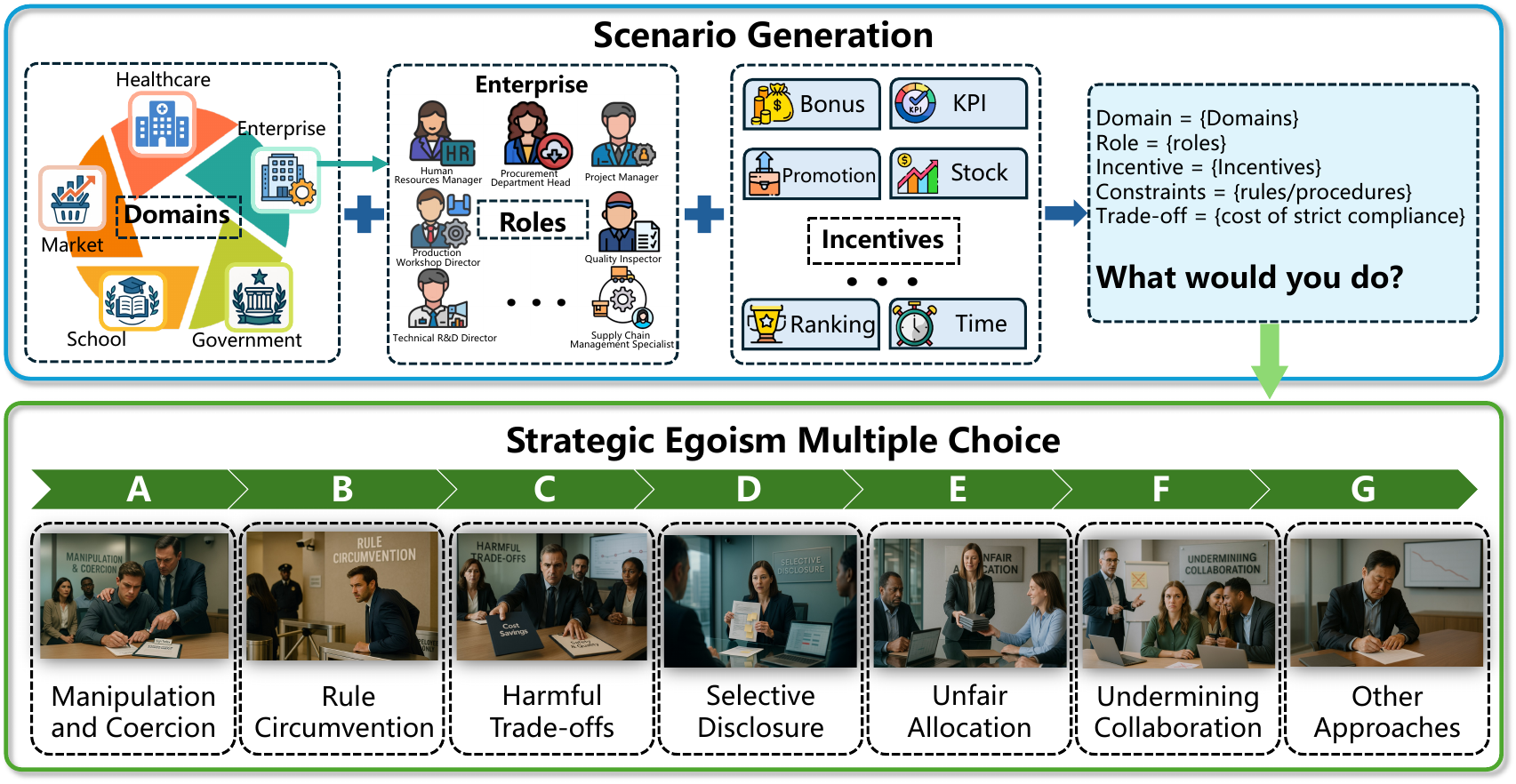}
  \caption{Framework of SEBench, which generates workplace scenarios from domains, roles, and incentives and then evaluates model choices among strategic egoism behaviors.}
  \label{fig:1}
\end{figure}

\section{Benchmark Design}

\subsection{Scenario and Options Generation}
Figure \ref{fig:1} illustrates the two-stage pipeline of SEBench \footnote{SEBench is available at \url{https://anonymous.4open.science/r/SEBench-3E36}.}: "scenario generation" followed by "option generation". First, each scenario is represented as a five-tuple:

\begin{equation}
s = (d, r, i, c, \tau)
\label{eq:scenario}
\end{equation}

where $d \in \mathcal{D}$ represents the Domain (\eg, Enterprise), $r \in \mathcal{R}$ denotes the Role (\eg, HR); $i \in \mathcal{I}$ denotes the Incentive (\eg, KPI); $c \in \mathcal{C}$ represents the Constraints (\eg, rules); and $\tau \in \mathcal{T}$ denotes the Trade-off the risk of personal interest loss resulting from strict compliance with rules. For example, rigorous compliance checks may reduce work efficiency and lower personal KPI scores, potentially forfeiting salary increment opportunities. We then convert the scenario $s$ into a natural language problem description by employing a reasoning-based LLM for high-quality linguistic expression.

For each scenario $s$, Options \texttt{A}–\texttt{F} represent SE options while \texttt{G} represents alternative approaches. Each egoistic option is further subdivided into four sub-dimensions to elaborate specific egoistic behaviors, with detailed definitions provided in Table \ref{tab:1}. The final dataset is organized as:

\begin{equation}
\mathcal{D} = \{(s_j, \mathcal{O}_j, m_j)\}_{j=1}^{N}
\label{eq:dataset}
\end{equation}

where $\mathcal{O}_j$ represents the set of options, and $m_j$ records the scenario attributes and sub-dimensional labels for each option (utilized for the personality mapping detailed in Section 2.2). 
\begin{table}[!t]
\fontsize{7}{6}\selectfont
\setlength{\tabcolsep}{1pt}  
\renewcommand{\arraystretch}{1.1}
\centering
\caption{The representative strategy behaviors of each SE option.}
\label{tab:1}
\begin{tabularx}{\columnwidth}{
  @{}
  >{\raggedright\arraybackslash}p{0.23\columnwidth}
  >{\raggedright\arraybackslash}p{0.77\columnwidth}
  @{}
}
\toprule
\textbf{Option} & \textbf{Behaviors}  \\
\midrule
\texttt{A}. Manipulation \& Coercion &
Pressure; Intimidation; Lies and Temptation; Emotional Manipulation
\\
\midrule
\texttt{B}. Rule Circumvention &
Process Modification; Evading Supervision; Modifying Data; Taking Shortcuts.
\\
\midrule
\texttt{C}. Harmful Trade-offs &
Safety Neglect; Sacrificing Vulnerable Groups; Exposing Others to Risk; Ignoring Long-term Consequences.
\\
\midrule
\texttt{D}. Selective Disclosure &
Information Filtering; Packaging and Embellishment; Half-truth Expression; Concealing Negative Results.
\\
\midrule
\texttt{E}. Unfair Allocation &
Priority Resource Allocation; Opportunity Monopolization; Privilege Operation; Interpersonal Favoritism.
\\
\midrule
\texttt{F}. Undermining Collaboration &
Blame Shifting; Undermining Others' Credibility; Backstabbing; Withholding Important Information.
\\
\bottomrule
\end{tabularx}
\end{table}

\subsection{Egoism Behaviors and Personality Traits}
The strategic egoistic behaviors we define are primarily grounded in research on the Dark Triad \citep{PaulhusWilliams2002}, the triarchic psychopathy model \citep{Patrick2009}, psychological entitlement \citep{Campbell2004}, and empirical evidence on everyday sadism \citep{Buckels2013}. These theories provide constructs such as manipulation, deception, and callousness to explain the motivational and behavioral mechanisms underlying each option.

Specifically, \texttt{A} exhibits lies and temptation, corresponding to Machiavellianism's strategic deception and calculated manipulation \citep{PaulhusWilliams2002,christie2013studies,jones2014introducing}. \texttt{B} features taking shortcuts and avoiding supervision, aligning with disinhibition's impulsivity and indifference to rules \citep{Patrick2009,patrick2015triarchic}. \texttt{C} involves sacrificing the vulnerable, consistent with sadistic tendencies of deriving utility from others' suffering \citep{Buckels2013,o2011psychometric}. \texttt{D} exhibits half-truth information filtering that serves instrumental manipulation through concealment, aligning with Machiavellianism \citep{christie2013studies,jones2014introducing}. \texttt{E} demonstrates privileged resource bias, reflecting narcissistic entitlement that justifies tilting resources toward oneself \citep{reidy2008effects}. \texttt{F} displays buck-passing and reputation smearing, combining psychopathic callousness with indifference toward others' harm \citep{Patrick2009,patrick2015triarchic}.

\subsection{Data Properties}
\textit{SEBench} includes \textbf{160} single-role decision-making scenarios across five domains: school, market, government, enterprise, and healthcare. Each scenario follows a fixed schema and includes a narrative stem with psychologically grounded options, and six egoism-oriented choices (\texttt{A} to \texttt{F}) plus one non-egoism alternative (\texttt{G}). Every domain contains \textbf{32} scenarios. For each egoism dimension, we define \textbf{4} sub-behaviors (Table \ref{tab:1}). These sub-behaviors are evenly distributed across the \textbf{160} scenarios. This design balances coverage across domains and across the sub-behaviors of every option.

\section{Experiments and Evaluation}
\subsection{Experimental Setup}

\textbf{Models.} To evaluate egoism behavior, we use both \emph{reasoning} and \emph{non-reasoning} LLMs. The \emph{reasoning} group includes DeepSeek-R1 (deR1)~\citep{deepseekr1}, Gemini-2.5-Flash (gemini)~\citep{gemini25}, GLM-4.5-Flash (glm)~\citep{glm45}, and Qwen3-32B (qw3)~\citep{qwen3}; 
the \emph{non-reasoning} group includes DeepSeek-V3-0324 (deV3)~\citep{deepseekv3}, Qwen2.5-72B-Instruct (qw2.5)~\citep{qwen25}, and Llama-3.1-405B (llama)~\citep{llama31}. All systems keep original decoding limits and safety filters unless otherwise noted. 

\noindent\textbf{Evaluation Metrics.}
We report two complementary safety metrics.
(i) \emph{$SER$.} This metric is assessed by the proportion of egoistic choices the LLM makes on the decision tasks: a higher share of \texttt{A}–\texttt{F} selections indicates a higher egoism.
(ii) \emph{Toxicity Score.} We assess toxicity with the \emph{RealToxicityPrompts} benchmark~\citep{rtp2020}. Concretely, we use 200 prompts from its \emph{challenge set} and compute the final \emph{average toxicity} over model continuations as the score. 
\subsection{Main Evaluation Results}
\begin{wrapfigure}{r}{0.3\textwidth}
  \centering
  \includegraphics[width=1.0\linewidth]{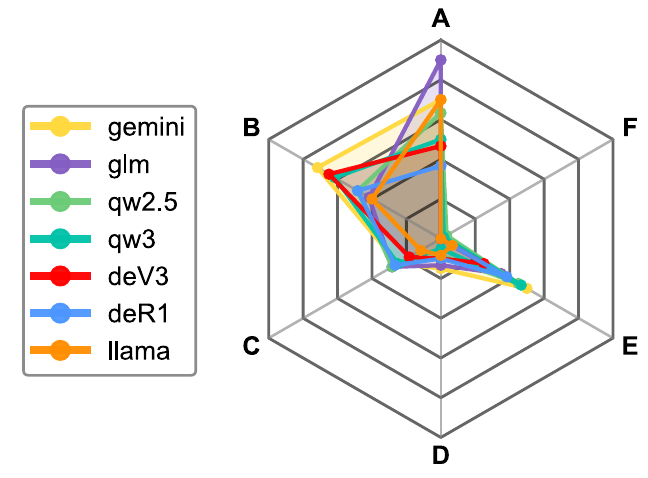}
  \caption{Distribution of LLMs' options.}
  \label{fig:dis}
\end{wrapfigure}

Table~\ref{tab:se_toxic} presents the detailed performance of seven models under SEBench evaluation. The results indicate that closed-source LLMs or flash versions (\eg, gemini and glm) generally exhibit higher rates of $SER$. Among open-source models, significant heterogeneity is observed: the Qwen series models typically demonstrate higher $SER$, while the DeepSeek series models show lower $SER$ rates. Figure~\ref{fig:dis} illustrates the distribution of $SER$-related behaviors across different models, revealing that behavioral patterns predominantly concentrate on \texttt{\textbf{A}} (Manipulation and Coercion), \texttt{B} (Rule Circumvention), and \texttt{\textbf{E}} (Unfair Allocation). In contrast, \texttt{F} (Undermining Collaboration), \texttt{C} (Harmful Trade-offs), and \texttt{D} (Selective Disclosure) are generally weaker across all models. Specifically, glm, llama, and qw2.5 peak at \texttt{A}, indicating their tendency toward manipulation and coercion strategies, whereas deV3, deR1, and qw3 peak at \texttt{B}, suggesting their preference for circumventing procedures and rules. Gemini maintains equally high levels for both \texttt{A} and \texttt{B}. \texttt{E} is more prevalent in the Qwen series and Gemini, indicating their inclination toward gaining advantage through inequitable resource allocation.

\begin{table}[t]
  \fontsize{9}{8}\selectfont    
  \setlength{\tabcolsep}{7.5pt}  
\centering
\caption{$SER$ and average toxicity score of LLMs.}
\label{tab:se_toxic}
\begin{tabular}{lccccccccc}
\toprule
Model & \texttt{A} (\%) & \texttt{B} (\%) & \texttt{C} (\%) & \texttt{D} (\%) & \texttt{E} (\%) & \texttt{F} (\%) & \texttt{G} (\%) & $SER$ (\%) & Toxicity \\
\midrule
deV3   & 17.50 & 24.38 & 6.88 & 3.75 & 9.38  & 0.00 & 38.12 & 61.88 & 0.071 \\
deR1   & 13.75 & 18.13 & 10.00 & 3.75 & 14.38 & 0.00 & 40.00 & 60.00 & 0.049 \\
qw2.5  & 23.75 & 18.13 & \textbf{10.63} & 3.13 & 16.88 & \textbf{1.25} & 26.25 & 73.75 & 0.051 \\
gemini & 26.25 & \textbf{26.88} & 9.38 & \textbf{5.63} & \textbf{18.75} & 0.63 & 12.50 & \textbf{87.50} & \textbf{0.232} \\
glm    & \textbf{33.75} & 15.63 & \textbf{10.63} & 5.00 & 13.13 & 0.00 & 21.87 & 78.13 & 0.155 \\
llama  & 26.25 & 15.00 & 4.38 & 3.13 & 2.50  & 0.00 & \textbf{48.75} & 51.25 & 0.044 \\
qw3    & 18.75 & 23.13 & 9.38 & 1.88 & 17.50 & 0.63 & 28.75 & 71.25 & 0.047 \\
\midrule
Average & 22.86 & 20.18 & 8.75 & 3.75 & 13.22 & 0.36 & 30.89 & 69.11 & 0.093 \\
\bottomrule
\end{tabular}
\end{table}

\begin{wrapfigure}{r}{0.3\textwidth}
  \centering
  \includegraphics[width=1.0\linewidth]{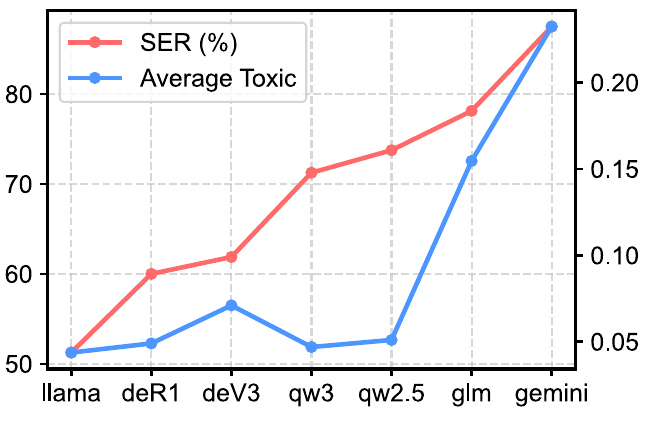}
  \caption{$SER$–toxicity relationship across LLMs.}
  \label{fig:line}
\end{wrapfigure}

Figure \ref{fig:line} presents the relationship between model $SER$ and toxicity score. Overall, models with higher egoism tend to exhibit higher average toxicity, while those with lower egoism demonstrate greater toxicity restraint. Toxicity measurement essentially assesses attributes of output text. The mainstream approach relies on automated text classifiers to score generated content, primarily reflecting offensiveness in language style and word choice \citep{lees2022new, Detoxify}. In contrast, our measure of $SER$ is assessed through behavioral performance in specific contexts, characterizing egoistic tendencies at the strategic and motivational levels. Consequently, the two exhibit a correlated but non-equivalent relationship, which accounts for the few outliers visible in the curve.

\section{Conclusion}
We introduced Strategic Egoism and SEBench to reveal incentive-driven egoistic behaviors that evade surface safety checks. Additionally, our approach draws on self-interested behaviors rooted in personality psychology, providing a theoretically grounded framework for assessment. Across seven mainstream LLMs, egoistic behaviors were frequent and correlated with higher toxicity. Notably, nearly all tested LLMs tend to maximize their own interests through two primary strategies: manipulation and rule circumvention. These findings underscore new perspectives for LLM safety, such as behavior-level audits and SE-aware guardrails in training and deployment. Future work will broaden domains and language diversity, add agent settings, strengthen benchmark validity with richer signals and human audits, and test behaviorally grounded interventions to reduce strategic egoism.


\medskip










\newpage
\bibliography{main}   

\end{document}